\documentclass[twocolumn,pre,showpacs]{revtex4}
\usepackage{graphicx}%
\usepackage{dcolumn}
\usepackage{amsmath}
\usepackage{latexsym}
\draft

\begin{document}

\def\bea{\begin{eqnarray}}
\def\eea{\end{eqnarray}}
\def\beq{\begin{equation}}
\def\eeq{\end{equation}}
\def\f{\frac}
\def\k{\kappa}
\def\sx{\sigma_{xx}}
\def\sy{\sigma_{yy}}
\def\sxy{\sigma_{xy}}
\def\e{\epsilon}
\def\ve{\varepsilon}
\def\ex{\epsilon_{xx}}
\def\ey{\epsilon_{yy}}
\def\exy{\epsilon_{xy}}
\def\be{\beta}
\def\D{\Delta}
\def\h{\theta}
\def\r{\rho}
\def\a{\alpha}
\def\s{\sigma}
\def\kb{k_B}
\def\la{\langle}
\def\ra{\rangle}
\def\nn{\nonumber}
\def\bu{{\bf u}}
\def\bn{\bar{n}}
\def\br{{\bf r}}
\def\up{\uparrow}
\def\dn{\downarrow}
\def\S{\Sigma}
\def\dg{\dagger}
\def\d{\delta}
\def\p{\partial}
\def\l{\lambda}
\def\G{\Gamma}
\def\o{\omega}
\def\g{\gamma}
\def\kv{\bar{k}}
\def\ha{\hat{A}}
\def\hv{\hat{V}}
\def\hg{\hat{g}}
\def\hG{\hat{G}}
\def\hTT{\hat{T}}
\def\noi{\noindent}
\def\a{\alpha}
\def\d{\delta}
\def\p{\partial} 
\def\nn{\nonumber}
\def\r{\rho}
\def\xv{\vec{x}}
\def\rv{\vec{r}}
\def\fv{\vec{f}}
\def\ov{\vec{0}}
\def\vv{\vec{v}}
\def\la{\langle}
\def\ra{\rangle}
\def\e{\epsilon}
\def\o{\omega}
\def\n{\eta}
\def\g{\gamma}
\def\th{\hat{t}}
\def\uh{\hat{u}}
\def\break#1{\pagebreak \vspace*{#1}}
\def\f{\frac}
\def\hf{\frac{1}{2}}
\def\uu{\vec{u}}

\bibliographystyle{/home/debc/Bibliography/prsty}
\title{Semiflexible polymers: Dependence on ensemble and boundary orientations} 
\author{Debasish Chaudhuri}
\affiliation
{Department of Biological Physics,
Max Planck Institute for the Physics of Complex Systems,
N{\"o}thnitzer Str. 38, 01187 Dresden, Germany
}
\email{debc@mpipks-dresden.mpg.de}

\date{\today}

\begin{abstract}
We show that the mechanical properties of a worm-like-chain (WLC) 
polymer, of contour length $L$ and persistence length $\lambda$ such that
$t=L/\lambda\sim{\cal O}(1)$,
depend both on the ensemble and the constraint on end-orientations.
In the Helmholtz ensemble, multiple minima in the free 
energy near $t=4$ persists for all kinds of orientational
boundary conditions. The qualitative features of projected 
probability distribution of end to end vector depend
crucially on the embedding dimensions.
A mapping of the WLC model, to a quantum particle moving on the
surface of an unit sphere, is used to obtain the
statistical and mechanical properties of the polymer under various 
boundary conditions and ensembles. The results show excellent
agreement with Monte-Carlo simulations. 

\end{abstract}
\pacs{87.15.La,36.20.Ey,87.15.Ya}
\maketitle

\section{Introduction}
\label{intro}
Microtubules and actin polymers constitute the structure of 
cytoskeleton that gives shape, strength and motility to most of the 
living cells. They are semiflexible polymers in the sense that 
their persistence lengths $\l$ are of the order of their chain 
lengths (the statistical contour lengths) $L$ 
such that the stiffness parameter $t=L/\l$ is small and finite. 
For example, actin, microtubule and double stranded DNA (dsDNA) 
have $\l=16.7~\mu m$\cite{ott,gittes},
$5.2~mm$\cite{gittes} and $50~nm$\cite{1dna} respectively. 
In physiological situation $L$ of dsDNA inside a cell may vary in 
between millimeters to a meter with average length in human being
$\sim 5 cm$, whereas typical contour lengths of 
microtubules can be $\sim 10 \mu m$\cite{cell}. The contour length
of actin filament can be as large as $100\mu m$\cite{ott}.
In the {\em in vitro} experiments, the contour lengths of bio-polymers
can be tailored chemically; e.g. in the experiment described in Ref.\cite{ott} 
the contour lengths of actin polymer have a distribution up to
$L=30\mu m$. For a polyelectrolyte like DNA
the persistence length $\l$ can also be tuned a little by changing
the salt concentration of the medium.
The relevant parameter in deciding the mechanical properties is the
stiffness parameter $t$, contour length measured in units of 
persistence length.
While it is obvious that in the thermodynamic limit of $t\to\infty$, 
the Gibbs (constant force) and the Helmholtz (constant extension) ensemble
predict identical properties, the same is not true for real semiflexible
polymers which are far away from this limit. 
In biological cells actin filaments remain dispersed throughout 
the cytoplasm with higher concentration in the cortex region, 
just beneath the plasma membrane. microtubules, on the other hand,
have one end attached to a microtubule-organizing centre,  centrosome,
in animal cells. Thus biologically important polymers may float freely
or may have one of their ends fixed. Even the end orientations 
of polymers play a crucial role in many important phenomena. For instance,
microtubule-associated proteins 
attach one or both of their ends to microtubules to arrange them in
microtubule bundles \cite{cell}. Again, in gene-regulation  
often DNA-binding proteins loop DNA with fixed end orientations
\cite{dna-loop-1,dna-loop-2,dna-loop-3}.
Thus it becomes important to understand the statistics and the mechanical
properties of semiflexible polymers with different possibilities
of end orientations and ensembles.

During the last decade many single molecule experiments have been 
performed on  semiflexible polymers\cite{smith,busta,smol,1dna}. 
These have been done by using the optical tweezers\cite{busta}, 
the magnetic tweezers\cite{magtwz} and the AFMs\cite{afm}.
In the optical tweezer experiments one end of a polymer is attached to
a dielectric bead which is, in turn, trapped by the light intensity profile 
of a laser tweezer. In this case the dielectric bead is free to rotate 
within the optical trap. On the other hand,
attaching an end of a polymer to a super-paramagnetic bead,
one can use magnetic field gradients to trap the polymer using a magnetic 
tweezer setup. In this case one can rotate the bead while holding it fixed
in position by changing the direction of the external magnetic field. 
In the AFM experiments one end of a polymer is trapped
by a functionalized tip of an AFM cantilever.
The two distinct procedures which can be followed to measure
force-extension are: 
(a) 
Both the ends of the polymer are held via the laser or magnetic tweezers or 
the AFMs. 
(b) 
One end of the polymer is attached to a substrate such that the position
and orientation of this end is fixed while the other end is trapped via a
laser or magnetic tweezer or an AFM cantilever. 

While the optical tweezers allow free rotation of dielectric beads 
within the trap, thereby, allowing free orientations of the polymer end,
the magnetic tweezers fix the orientation of the ends and one can study the 
dependence of polymer properties on end-orientations by controlled 
change of the direction of external magnetic field. 
In this paper, we call this fixing of orientation of an end of 
a polymer as grafting. By changing the trapping potential 
from stiff to soft trap one can go from the Helmholtz to the Gibbs 
ensemble\cite{sam-ensemble}. 
Before we proceed, let us first elaborate on how to fix the
ensemble of a mechanical measurement\cite{sam-ensemble,kreu}. 
In the simplest case we can assume 
that one end of the polymer is trapped in a harmonic well,
$ V(z) = C (z-z_0)^2/2$
with ($0,0,z_0$) being the position of the potential-minimum. 
The polymer end will undergo continuous thermal motion. One can
use a feedback circuit to shift $z_0$
to force back the fluctuating polymer end to its original position.
This will ensure a Helmholtz ensemble.
This can also be achieved by taking $C\to\infty$.
On the other hand, one can use a feedback circuit to fix the force 
$-C(z-z_0)$ by varying $C$ depending on the position $z$ of the
polymer end. This will ensure a Gibbs ensemble.
This can also be achieved by taking a vanishingly soft
($C\to 0$) trap to infinitely large distance ($z_0\to\infty$)
such that within the length scale of fluctuation the polymer end 
feels a constant slope of the parabolic potential. Surely, in experiments,
using a feedback circuit is easier to implement a particular ensemble.
However, the other procedure is mathematically well defined and one
can seek recourse of it to show that the partition function
of two ensembles are related by a Laplace transform~\cite{sam}.
This relation does not depend on the choice of the Hamiltonian 
for a polymer.
An exact relation between the two ensembles for worm like
chain (WLC) model is shown in Sec.\ref{pathint}.

From the above discussion on possible experiments, it is clear 
that there can be three possibilities of boundary conditions 
in terms of orientations:
(a) Free end: Both the ends of a polymer can remain free to 
rotate\cite{mywlc,sam}.
(b) One end grafted: Orientation of one end is fixed and the 
other can take all possible orientations\cite{munk-frey}.  
(c) Both ends grafted: Orientations of both the ends are kept fixed.
Thus, in experiments, one can have two possible ensembles and 
three possible boundary conditions. 
We restrict ourselves to the WLC polymers embedded in two dimensions (2D).
We investigate the probability distribution, free 
energy profile and force extension relation for each of 
these cases in this paper. We shall see that the properties
of a semiflexible polymer depend both on the choice of the ensemble and
the boundary condition. Note that, there can be other possibilities of boundary
conditions e.g. orientation at one end of a polymer can be free to rotate 
on a half-sphere\cite{pnelson}. However, in this paper we focus
on the three  possible boundary orientations listed above.

The WLC model is a simple coarse grained way to capture
bending rigidity of an unstretchable polymer~\cite{kratky,doi}
embedded in a thermal environment. 
Recent single molecule experiments in biological
physics~\cite{smith,busta,smol,1dna} renewed interest in this old model
of polymer physics. It was successfully employed ~\cite{bmss,marko} 
to model data of force-extension experiments~\cite{smith} on dsDNA.
Mechanical properties of giant muscle protein titin~\cite{titin,sac-tit},
polysaccharide dextrane~\cite{afm,sac-tit} and single molecule of 
xanthane~\cite{xan} were also explained using the WLC model. 
Due to the inextensibility constraint, the WLC model is hard to tract
analytically except for in the two limits of flexible chain ($t\to\infty$) 
and rigid rod ($t\to 0$), about which perturbative calculations have been 
done~\cite{dan,gob,nori,chervy}. A key quantity that describes statistical
property of such polymers is the distribution of end-to-end separation.
Numerical simulations to obtain radial distribution function for
different values of $t$ have been reported along with a series 
expansion valid in the small $t$ limit\cite{wilh}. 
Mean-field treatments to incorporate the inextensibilty in an approximate
way have also been reported\cite{thiru,jkb}. 
In an earlier study\cite{mywlc} we investigated the free energy profile of 
a semiflexible polymer whose ends were free to rotate in the 
constant extension ensemble and in the stiffness regime of 
$1\le t\le 10$. This work predicted that a clear qualitative signature of
semiflexibility would be a non-monotonic force extension 
for stiffnesses around $t\sim 4$ in the Helmholtz 
ensemble. This comes from the multimodality of probability
distribution of end to end separation. However, this non-monotonicity 
is absent in the Gibbs ensemble\cite{mywlc}. 
Multiple maxima in the probability distribution of end to end separation
was due to a competition between entropy, that prefers a maximum near 
zero separation, and energy, that likes an extended polymer.
A series of later studies \cite{sam,kleinert-1,kleinert-2,stepanow} 
used analytic techniques to understand the end to end distribution
at all stiffnesses including the stiffness regime where multimodality
was observed.
Recently, multimodality is found 
in transverse fluctuations of a grafted polymer using simulations
\cite{munk-frey} and approximate theory\cite{kroy,munk-frey-th}.
A Greens function technique has been developed
that takes into account the orientations of the polymer ends \cite{wang}.
The impact of the specific boundary conditions and the comparable length
scales of a dsDNA and the beads to which it is attached in typical
force-extension measurements have been identified in another recent 
study~\cite{pnelson}.
The WLC model has also been extended to study statistics of end to end 
separation and loop formation probability in
dsDNA\cite{sunil} and to incorporate twist 
degree of freedom~\cite{twist-1,twist-2,twist-3,twist-4}. 

The construction of this paper is as follows. 
In Sec.\ref{pathint} we present a theoretical technique
for exact calculation of the WLC model via a mapping to a quantum particle
moving on the surface of an unit sphere. This technique incorporates
all the possible end orientations and predicts results in both the
Helmholtz and the Gibbs ensembles. In Sec.\ref{simu} we discuss 
the different discretized versions of the WLC model and the Monte-Carlo 
(MC) simulation procedures followed in this work.
In Sec.\ref{result} we present all the results of probability 
distributions and force-extensions etc. obtained from
theory and simulations. 
Then, in Sec.\ref{conclusion}, we summarize our results and
conclude with some discussions.

\section{Theory}
\label{pathint}

In the WLC model a polymer is taken as a continuous curve denoted
by a $d$-dimensional vector $\rv(s)$ where $s$ is a distance measured over the 
contour of the curve from one of its ends. This curve has a bending 
rigidity and thus the Hamiltonian is given by
\bea
\be{\cal H} &=& \f{\k}{2} \int_0^L ds~ \left(\f{\p\th(s)}{\p s}\right)^2,
\label{eqwlc}
\eea
where $\th(s)=\p\rv(s)/\p s$ is the tangent vector and the polymer is
inextensible i.e. $\th^2=1$, $\be$ is the inverse temperature. 
Persistence length is a
measure of the distance up to which the consecutive tangent vectors on
the contour do not bend appreciably and is defined by
$\la \th(s).\th(0)\ra = \exp(-s/\l)$.
The bending rigidity $\k$ is  related to
persistence length $\l$ via $\k = (d-1)\l/2$. 

In this section we present a theoretical method to solve the WLC model to
any desired accuracy\cite{saito,sam} 
for both the Helmholtz and the Gibbs ensembles and all the three 
possible boundary orientations over the entire range of stiffness parameter $t$.
We first present the method for a free polymer\cite{sam}. Then we extend
it to calculate properties of grafted [~one/both end(s)~] polymers.

The partition function of a WLC polymer in the Helmholtz ensemble is 
$Z(\rv)=\sum_{ c}\exp(-\be{\cal H})$ where ${ c}$ denotes a sum
over  all possible configurations of
the polymer that are consistent with the inextensibility constraint. 
The probability distribution of the end to end vector becomes,
$P(\rv) = Z(\rv)/\int^L d\rv Z(\rv) = {\cal N} Z(\rv)$.
If the tangent vectors of the two ends of a polymer are held fixed at 
$\th_i$ and $\th_f$, the probability distribution of end to end vector 
in constant extension ensemble can be written in path integral notation as
\bea
P(\rv) = {\cal N} \int_{\th_i}^{\th_f} {\cal D}[\th(s)] \exp\left(-\be{\cal H}\right) 
\times \d^d\left( \rv - \int_0^L \th ds\right)
\eea
where 
${\cal D}[\th(s)]$ denotes integration over all possible paths in
tangent vector space from the tangent at one end $\th_i$ to the tangent 
at the other end $\th_f$. In $d$-dimensions $\rv=(r_1,r_2,\dots,r_d)$.
Recently
a path integral Greens function formulation has been developed \cite{sam}
to evaluate the end to end distribution for a free
polymer in 3D. We closely follow that method and generalize it to obtain
results for various orientation constraints on polymer ends.
In particular we focus on polymers living in a 2D embedding space.

The integrated (projected) probability distribution is given by,
\bea
P_x(x) = \int d\rv P(\rv)\d (r_1 -x).
\eea
We define the generating function of $P_x(x)$ via a Laplace transform,
\bea
\tilde P(f) 
=\int_{-L}^L dx \exp(fx/\l) P_x(x)
\label{zf}
\eea
where $f$ is the force in units of $\kb T/\l$ i.e. $f=F\l/\kb T$ applied
along the $x$-axis. Again, the partition function in the Gibbs ensemble,
$\tilde Z(\fv) = \int^L d\rv \exp(\fv.\rv/\l) Z(\rv)$\cite{sam}. 
This immediately gives, ${\cal N}=1/\tilde Z(\ov)$. 
We show that $\tilde Z(\ov)$ is a constant which 
depends on the constraints on end orientations.
Eq.\ref{zf} gives,
\bea
\tilde P(f) &=& {\cal N} \int_{\th_i}^{\th_f} {\cal D}[\th(s)] e^{\left(
-\f{(d-1)\l}{4}\int_0^L ds \left( \f{\p \th (s)}{\p s}\right)^2 
+ \f{f}{\l}\int_0^L \th_x ds\right)} \nn\\
&=& {\cal N} \int_{\th_i}^{\th_f} {\cal D}[\th(\tau')] e^{\left[
-\int_0^{t} \left\{ \f{(d-1)}{4} \left(\f{\p \th (\tau')}{\p \tau'}\right)^2 
- f \th_x \right\} d\tau' \right]}\nn\\
\eea
The last step is obtained by replacing $\tau' = s/\l$ and
using the identities $\k = (d-1)\l/2$ and $t=L/\l$.
Note that, $\tilde P(f)$, is the partition function, apart from a 
multiplicative constant, in the Gibbs ensemble where $t$ behaves 
like an inverse temperature such that the Gibbs free energy can be written 
as $G(f) = - 1/t~ \ln \tilde P(f)$. 
Now considering $\tau'$ as imaginary time and replacing $\tau = - i \tau'$ 
one gets,
\bea
\tilde P(f) 
={\cal N} \int_{\th_i}^{\th_f} {\cal D}[\th(\tau)] e^{ \left[i \int_0^{-it} {\cal L} d\tau \right]}~~;
\eea
with the identification of  
${\cal L}= \f{(d-1)}{4} \left(\f{\p \th (\tau)}{\p \tau}\right)^2 + f \th_x$ 
as the Lagrangian, $\tilde P(f)$
[~= $\tilde Z(f)/\tilde Z(0)$~] in the above expression 
is the path integral representation for the propagator
of a {\em quantum} particle, on the surface of a $d$- dimensional sphere,
that takes a state $|\th_i\rangle$ to $|\th_f\rangle$. 
In Schrodinger picture this can be written
as the inner product of a state $|\th_i\rangle$ and another state 
$|\th_f\rangle$ evolved by imaginary time $-it$, 
\bea
\tilde Z(f) = \langle \th_i| \exp(-i \hat H (-it))|\th_f\rangle = 
\langle \th_i| \exp(-t \hat H )|\th_f\rangle ,
\label{prop}
\eea
where $\hat H$ is the Hamiltonian operator corresponding to the
Lagrangian ${\cal L}$.

Once $\tilde P(f)=\tilde Z(f)/\tilde Z(0)$ is calculated, 
performing an inverse Laplace transform one can 
obtain the projected probability distribution $P_x(x)$. 
Eq.\ref{zf} can be written as,   
\bea
\tilde P(f) 
=\int_{-1}^1 dv_x \exp(tf~v_x) p_x(v_x)
\label{pf}
\eea
where $v_x = x/L$ and $p_x(v_x)=LP_x(x)$ is a scaling relation.  
Note that the Helmholtz free energy is given by 
${\cal F}_x(v_x)=-(1/t)\ln p_x(v_x)$. 
Thus Eq.\ref{pf} gives the relation between the Helmholtz and
the Gibbs ensemble for finite chain (finite $t$),
\bea
\exp[-tG(f)]=\int_{-1}^1 dv_x \exp(tf~v_x) \exp[-t{\cal F}_x(v_x)].\nn
\eea
In thermodynamic limit of $t\to\infty$,
a steepest descent approximation of the above integral relation gives 
$G(f)={\cal F}_x(v_x)-fv_x$, the well known Legendre transform relation.
Identifying $-iu = tf$ one can define Fourier transform relations,
$\tilde p_x(u) = \int_{-1}^1 p_x(v_x) \exp(-iuv_x) dv_x$
and 
\bea
p_x(v_x) = \f{1}{2\pi} \int_{-\infty}^\infty du  \tilde p_x(u) \exp(iuv_x) 
\label{nlap}
\eea
such that
$\tilde P(f) = \tilde p_x(u=ift) $ and the inverse Fourier transform can be
written as an inverse Laplace transform,
\bea
p_x(v_x) = t \f{1}{2\pi i}\int_{-i\infty}^{i\infty} df 
\tilde P(f) \exp(-tfv_x) ~~.
\eea
The simplest way to obtain $p_x(v_x)$, numerically, is 
to replace $f=-iu/t$ in the
expression for $\tilde P(f)$ to obtain $\tilde p_x(u)$ 
and evaluate the inverse Fourier transform (Eq.\ref{nlap}).

\begin{figure}[t]
\begin{center}
\includegraphics[width=8cm]{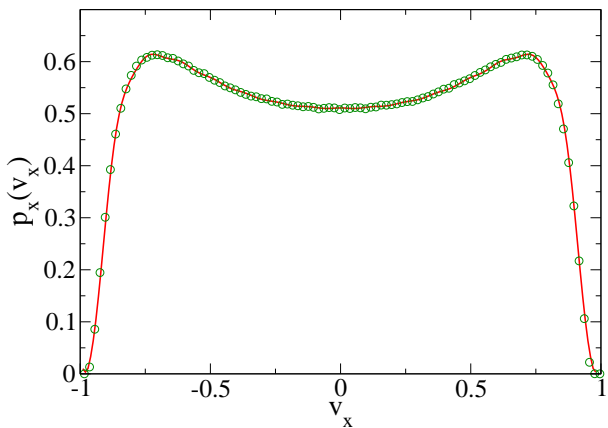}
\end{center}
\caption{(Color online) 
For a semiflexible polymer in 2D having its ends free to rotate 
$p_x(v_x)~(~=p_y(v_y)~)$ is plotted at stiffness parameter $t=2$.  
The points are collected from Monte-Carlo simulation
in freely rotating chain model (see Sec.\ref{simu}). The line is calculated
from theory (see Sec.\ref{pathint}). The theory shows excellent agreement with 
simulation.  It clearly shows bimodality via two maxima in 
integrated probability distribution at the two near complete extensions.
}
\label{t2-fx-free}
\end{figure}

Up to this point everything has been treated in $d$- embedding dimensions. 
Experiments on single polymer can be performed in three dimensions (3D) 
as well as in two dimensions (2D).
In 3D, polymers are left inside a solution whereas one can float the polymer
on a liquid film to measure its properties in 2D \cite{gittes}.
However, polymers embedded in 2D are more interesting because of the 
following reason.
In a free polymer whose end orientations are free to rotate, the system
is spherically symmetric and thus the probability distribution of end to
end vector $P(\vec r)=P(r)$ where $r=|\vec r|$. For this system 
it was shown that in the Helmholtz ensemble in 3D\cite{sam},
\bea
p(v_x) = -\f{1}{2\pi v_x}~\f{dp_x}{dv_x}
\eea
where $p(v=r/L)=L^d P(r)$ is the probability distribution of end to
end distance scaled by the contour length $L$. In presence of the 
spherical symmetry of a free WLC polymer, this distribution gives 
the Helmholtz free energy ${\cal F}(v)=-(1/t)\ln p(v)$\cite{mywlc}.
$P(r)$ is related to the radial distribution function $S(r)$ via
$S(r)=C_d r^{d-1} P(r)$ where $C_d$ is the area of a $d$- dimensional 
unit sphere.
Since $p(v)$ is a probability distribution,
$p(v_x)\geq 0$ and therefore $dp_x/dv_x\leq 0$ for $v_x>0$ thus ruling
out multiple peaks in $p_x(v_x)$ \cite{sam-ensemble} and showing that 
$p_x(v_x)$ will have a single maximum at $v_x=0$ for all values of 
stiffness parameter $t$.
No such simple relation exists between $p(v_x)$ and $p_x(v_x)$ in 2D. 
The two dimensional
WLC polymer having its ends free to rotate may show more than one maximum
in $p_x(v_x)$ and therefore non-monotonicity in force-extension. 
Indeed our calculation
and simulation (see Sec.\ref{simu}) does show multiple maxima in
projected distribution $p_x(v_x)$ (Fig.\ref{t2-fx-free}). 
This is a curious difference between semiflexible 
polymers in 2D and 3D. Because of this and the fact that experiments in
2D are possible\cite{ott}, in this work we focus on the 2D WLC polymers.

We have already given a general form of $\tilde Z(f)$ (Eq.\ref{prop})
which depends on the dimensionality $d$ of the embedding space.
For $d=2$, one can
assume $\th = (\cos\h,\sin\h)$, leading to
${\cal L} = \{1/4~\dot\h^2 + f cos\h\}$.  This  automatically maintains
the inextensibility constraint $\th^2=1$.
The angular momentum $p_\h = \f{\p L}{\p \dot\h} = \dot\h/2$ and
thus the corresponding Hamiltonian  
$H = \dot\h p_\h - {\cal L} = p_\h^2 - f cos\h$. 
In planar polar coordinates, replacing $p_\h \to -i\f{\p}{\p \h}$ 
one obtains the corresponding quantum Hamiltonian
operator, $\hat H = - \f{\p^2}{\p \h^2} - f cos\h$. In this 
representation of tangent vectors, 
\bea
\tilde Z(f) &=& \langle \h_i| \exp(-t \hat H )|\h_f\rangle \nn\\
&=& \sum_{n,n'} \phi_n^\ast(\h_i) \phi_{n'}(\h_f) \langle n| \exp(-t \hat H )| n'\rangle~,
\label{zsym0}
\eea
where $\phi_n(\h)=\la n|\h\ra$.
%
If external force is applied along $x$- direction as in Eq.\ref{zf},
$\hat H = \hat H_0 + \hat H_I = 
- \f{\p^2}{\p \h^2} - f cos\h$. 
Thus the total Hamiltonian 
$\hat H$ denotes a rigid rotor ($\hat H_0= -\f{\p^2}{\p \h^2}$)
in presence of a constant external field ($\hat H_I= -f \cos\h$). 
The eigenvalues of $ \hat H_0 $ 
are $E_n = n^2$ and 
the complete set of orthonormal eigenfunctions are given by 
$\phi_n(\h) = \exp(i n \h)/\sqrt{2\pi}$ where 
$n=0,\pm 1,\pm 2,\dots,\pm\infty$. 
In this basis 
$\langle n|\hat H_I| n'\rangle = -(f/2)(\d_{n',n+1} + \d_{n',n-1})$. 
Therefore, 
$\langle n|\hat H| n'\rangle = n^2 \d_{n',n} - (f/2)(\d_{n',n+1} + \d_{n',n-1})$.
If the external force were applied in $y$- direction $\hat H_I= -f \sin\h$ and
$\langle n|\hat H| n'\rangle = n^2 \d_{n',n}-(f/2i)(\d_{n',n+1} - \d_{n',n-1})$.
$\langle n|\exp(-t\hat H)| n'\rangle$ can be calculated by exponentiating the
matrix $\langle n|\hat H| n'\rangle$. Thus one can find $\tilde Z(f)$ and
hence $\tilde P(f)$ and $p_x(v_x)$.

Note that the above formalism can be easily extended to find the end to end
vector probability distribution $p(v_x,v_y)$. A Laplace transform of $P(\rv)$
is $\tilde P(\fv)=\int^L d\rv \exp(\rv.\fv/\l)P(\rv)$. 
In a similar manner as above one can show that 
$\tilde P(\fv) = \tilde Z(\fv)/\tilde Z(\ov)$ with $\tilde Z(\fv)$
given by Eq.\ref{zsym0} with 
$\hat H = - \f{\p^2}{\p \h^2} - f_x \cos\h -f_y \sin\h$.
Thus, using an inverse Laplace transform one can find $P(\rv)$ and hence
$p(v_x,v_y)$.

\subsection{ Free polymer}  
For a polymer which has both its ends free to rotate,
integrating Eq.\ref{zsym0} over all possible initial and final tangent vectors
in rigid rotor basis one gets,
$\tilde Z(f) = 2\pi~\langle 0| \exp(-t \hat H )| 0\rangle$,
$\tilde Z(0) 
=2\pi$ and  hence
\bea
\tilde P(f)
= \langle 0| \exp(-t \hat H )| 0\rangle~.
\eea
This means that $\tilde P(f)$ is given by the $(0,0)$-th element of the matrix
$\langle n| \exp(-t \hat H )| n'\rangle$.
Thus, if the external force $f$ is applied in $x$- direction, 
remembering $\tilde p_x(u)=\tilde P(f=-iu/t)$ one can calculate 
the inverse Fourier transform (Eq.\ref{nlap}) to obtain $p_x(v_x)$.
In this case, due to spherical symmetry of a polymer whose 
ends are free to rotate, $p_x(v_x)=p_y(v_y)$. 

\subsection{One end grafted}
This symmetry breaks down immediately if one end of the polymer is fixed to a
specific direction, namely along the $x$-axis {\em i.e.} $\h_i = 0$. 
Then in Eq.(\ref{zsym0}) integrating over all possible $\h_f$  and 
leaving $\h_i=0$ one obtains
$\tilde Z(f) = \sum_{n}  \langle n| \exp(-t \hat H )| 0\rangle$ 
in the rigid- rotor basis. 
Note, for this case $\tilde Z(0) 
=1$ and therefore  
\bea
\tilde P(f)=\sum_{n}  \langle n| \exp(-t \hat H )| 0\rangle~.
\eea

\subsection{Both ends grafted}
Two ends of a polymer can be grafted in infinitely different ways.
Let us fix the orientation of one end along $x$- direction ($\h_i=0$)
and the other end 
along any direction $\h_f$. 
Then Eq.(\ref{zsym0}) gives
$2\pi\tilde Z(f) = \sum_{n,n'} e^{in'\h_f} \langle n| \exp(-t \hat H )| n'\rangle$,
$2\pi\tilde Z(0) 
= \sum_{n} e^{in\h_f-t n^2}$ and hence
\bea
\tilde P(f)=\f{\sum_{n,n'}e^{in'\h_f}\langle n| e^{-t \hat H }| n'\rangle}{\sum_{n} e^{in\h_f-t n^2}}~.
\eea

\vskip .5cm
If the external force is in  $x$- direction, 
the Laplace transform of $\tilde Z(f)$, defined in the way described above, 
gives the projected probability distribution in $x$- direction, $p_x(v_x)$. 
On the other hand, if the external force is in  $y$- direction, 
the Laplace transform of $\tilde Z(f)$ gives the projected probability 
distribution in $y$- direction $p_y(v_y)$, the distribution of transverse 
fluctuation while one end of the polymer is grafted in $x$- direction.

All the relations derived so far are exact. Since the calculation 
of an infinite order matrix $\la n|\exp(-t\hat H)|n' \ra$ is not feasible, 
we calculate it numerically\cite{matexp} by truncating 
up to an order $N_d$, that controls the accuracy, limited
only by computational power. 
Unless otherwise stated, we use $N_d=11$ which already gives 
very good agreement with simulated data (see Fig.\ref{t2-fx-free} and
Sec.\ref{result}). The inverse Laplace transforms 
to obtain end to end probability distributions from $\tilde P(f)$s are
also done numerically.

\section{Simulation}
\label{simu}
In this section, we introduce two discretized models that we
use to simulate semiflexible polymers. Both of these are
derived from the WLC model which has been used for our theoretical
treatment in Sec.\ref{pathint}. After introducing the discretized 
models we show how to impose the various boundary conditions on 
end orientations. We 
perform Monte-Carlo (MC) simulations of these models to obtain probability 
distributions in the Helmholtz ensemble. 

One discretized version of the Fokker-Plank equation corresponding to
the WLC model is the freely rotating chain (FRC) model\cite{dan,gob}. 
In the FRC model, one considers a polymer as a random walk
of $N$ steps each of length $b=L/N$ with one step memory, such that, successive
steps are constrained to be at an fixed angle $\h$ with $\l=2b/\h^2$. 
The continuum WLC model is obtained in the limit $\h,~b\to 0$, $N\to \infty$
keeping $\l$ and $L$ finite. To simulate a polymer with ends free to rotate
a large number of configurations are generated with first step taken in any
random direction. Whereas if one chooses the first step to be in some specific
direction, this will simulate a polymer with one end grafted in that 
direction.

A straight forward discretization of the Hamiltonian in 
Eq.\ref{eqwlc} in 3D (2D) 
is an 1d Heisenberg (classical XY) model:
\bea
\be{\cal H} &=& \f{\k}{2} \sum_{i=1}^N \f{(\th_i-\th_{i-1})^2}{b} 
= \sum_{i=1}^N (-J~\th_i.\th_{i-1})
\eea
with a nearest neighbor coupling $J=\k /b$ between `spins' $\th_i$.
We have ignored a constant term in energy. 
The appropriate continuum limit is recovered for $b\to 0$, $J\to \infty$ with
$Jb=\k$ finite. In this model grafting is simulated by fixing end spins on the
1D chain. If an end is free then the end spin takes up any orientation
that are allowed by the energy and entropy. In this model, by fixing the
two end-spins, one can easily simulate a polymer with both its ends
grafted in some fixed orientations. We follow the normal Metropolis 
algorithm\cite{frenkel} to perform MC simulation in this model.

We restrict ourselves to two dimensions.
In the FRC model 
simulations we have used a chain length of $N=10^3$ and generated around $10^8$
configurations. This simulation does not require equilibration run. 
Therefore all the $10^8$ configurations were used for data collection.
In the XY model we have simulated $N=50$ spins and equilibrated over $10^6$
MC steps. A further $10^6$ configurations were generated 
to collect data. We have averaged over $10^3$ initial
configurations, each of which were randomly chosen from 
nearly minimum energy configurations that conform with the boundary
conditions. Increasing $N$, in both the models of simulation, do not 
change the averaged data. As a check on the numerics, we compared
simulation evaluation of $\la r^2\ra$ and $\la r^4\ra$ with their
exact results\cite{mywlc,wang} to obtain agreement within around $0.5\%$.

Notice that in simulating the FRC model one performs random walk with
fixed angle between consecutive steps and does not require to equilibriate.
Thus one uses all the simulated configurations for data collection. On
the other hand, in simulating XY-model one has to perform equilibration
runs over a large number of steps and an averaging over many initial
configurations is required. Another important difference between the two
simulation methods is that, in the XY-model simulation, in each MC step one
has to calculate a time consuming exponential of change in energy, whereas,
no such exponential calculation is required in simulating the FRC model.
Thus simulating the FRC model is clearly much faster, computationally.
However, implementing the fixed boundary orientations at both the
ends of a polymer is much easier in the XY-model. 

\section{Results}
\label{result}
Once all these theoretical and simulation tools are available, 
we apply them to bring out the statistical and mechanical properties 
of a semiflexible polymer. We have three different boundary conditions
depending on the orientational constraints on the polymer ends and 
two different ensembles.
For each case we look at the various probability densities, ensemble 
dependence of force-extension etc. For the case of a polymer with both 
ends grafted we find that the properties depend on the relative 
orientation of the two ends.

\subsection{Free polymer}

\begin{figure}[t]
\begin{center}
\includegraphics[width=8.cm]{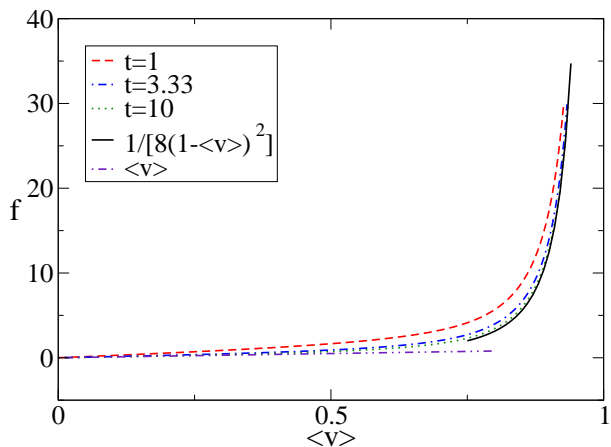}
\end{center}
\caption{(Color online)
None of the force-extension curves, obtained in the Gibbs ensemble, 
including that at $t=3.33$ show non- monotonic behavior unlike in the Helmholtz 
ensemble \cite{mywlc}. Forces are expressed in units of $\kb T/\l$ i.e.
$f=F\l/\kb T$.
}
\label{fx-free}
\end{figure}

{\em The Helmholtz ensemble:}
We employ the theory as described in Sec.\ref{pathint} to calculate
$p_x(v_x)$ and $p_y(v_y)$ for a polymer with both its ends free to rotate.
We compare the probability distributions obtained at stiffness parameter
$t=2$ with that obtained from MC simulation (Sec.\ref{simu}) using the 
FRC model (see Fig.\ref{t2-fx-free}). This shows excellent
agreement between theory and simulation. For a free polymer $p_x(v_x)$
and $p_y(v_y)$ are same due to the spherical symmetry.
Note that ${\cal F}(v_x) = -(1/t) \ln p_x(v_x)$ would give a 
non- monotonic force-extension $\la f_x\ra$-$v_x$ due to the multimodality in 
$p_x(v_x)$ (Fig.\ref{t2-fx-free}) via $\la f_x\ra = (\p {\cal F}/\p v_x)$. 
The force-extension obtained from the projected probability distribution 
$p_x(v_x)$
corresponds to the experimental scenario in which the external potential
traps the polymer end only in the $x$- direction and the polymer-end is
free in $y$.
In general, if the external potential traps the polymer-end in $d_r$ dimensions
($d_r\leq d$) then a $d_r$ dimensional projection
(~[$d-d_r$] dimensional integration~) of the probability distribution
of end to end vector $p(\vv)$ gives the appropriate free energy and decides the
force-extension relation. On the other hand, if the trapping potential
holds a polymer-end in all the $d$-dimensions, as is usually done
in most force-extension experiments, only the end to end vector distribution
$p(\vv)$ gives the appropriate Helmholtz free energy that can predict the
force-extension behavior in the Helmholtz ensemble.
This understanding is general and does not depend on the specific 
orientational boundary conditions or the dimensionality $d$ of
embedding space. This is important to keep in mind while analyzing
experimental data. In experiments that use the laser tweezers to trap
polymer ends in $d$- dimensions, ends remain free to rotate and the
relevant Helmholtz free energy is obtained from $p(v)$. 
Ref.\cite{mywlc} predicted multiple minima in this
free energy leading to non-monotonic force-extension in such experiments.

{\em The Gibbs ensemble:}
We have already mentioned that the non- 
monotonic nature of force-extension, a strong qualitative signature of 
semiflexibility, is observable only in the Helmholtz ensemble and 
not in the Gibbs ensemble \cite{mywlc}. In the Gibbs ensemble, the averaged
extension comes out to be $\la v \ra= -(\p G/\p f)$ and the response
$\p \la v \ra/\p f = t[\la v^2\ra - \la v\ra^2] \geq 0$.
Similar relation for response function does not exist in the Helmholtz ensemble.
Therefore, the force-extension in the Gibbs ensemble has to be
monotonic (Fig.\ref{fx-free}) in contrast to the Helmholtz ensemble.
For a polymer with its ends free to rotate, the force extension 
relations, that have been calculated from theory,  at various $t$ 
are shown in Fig.\ref{fx-free}.
For small forces the polymer shows linear response.
At large and positive force polymer goes to fully extended limit 
beyond which, the inextensibility constraint prevents further extention.
It is possible to do perturbative analysis of 
$\tilde P(f)=\la 0 |\exp(-t\hat H)|0\ra$ in the two extreme limits of small 
and large forces to obtain the asymptotic force-extensions\cite{ashok}. 
In the small force limit, $f \cos\h$ may be treated
as a perturbation about the rigid rotor hamiltonian $\hat H_0=-\p^2/\p\h^2$.
Thus keeping upto the second order correction to eigen-values we obtain 
$E_0 = - f^2/2$. Within this perturbative approximation 
$\tilde P(f)=\exp(-t E_0)$ and therefore $G(f)=-1/t~\ln \tilde P(f)= -f^2/2$.
Thus the force extension relation in this limit is 
$\la v\ra = -\p G/\p f =f$. 
On the other hand, for large forces one can expand the term
$\cos\h\simeq 1-\h^2/2$ and write $\hat H = -t f + t \hat H_0$ where
$\hat H_0 = -\p^2/\p\h^2+(1/2)f\h^2$ is the harmonic oscillator Hamiltonian.
In the harmonic oscillator basis, the ground state eigenvalue 
$E_0 = \sqrt{f/2} $ and thus the ground state energy corresponding to
$\hat H$ is $ -f +  \sqrt{f/2} $. Therefore, in a similar manner as in
above, $\tilde P(f)=\exp( -t\sqrt{f/2} + t f )$ and $G(f) = \sqrt{f/2} -f$.
Thus, for large forces, the force-extension relation comes out to be
$\la v\ra = -\p G/\p f = 1-1/(2\sqrt{2f})$ 
which can be inverted to get the relation,
$f=1/[8(1-\la v \ra)^2]$. All the curves $f(\la v \ra)$ in Fig.\ref{fx-free}
falls on to $f=\la v\ra$ at $f\to 0$ limit and to $1/[8(1-\la v \ra)^2]$ in
the $f\to\infty$ limit.

\begin{figure}[t]
\begin{center}
\includegraphics[width=8.cm]{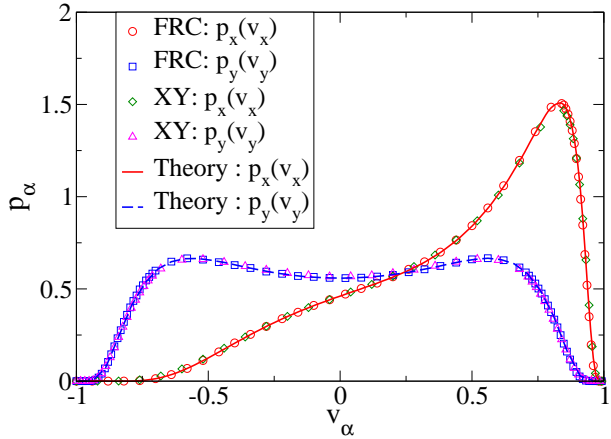}
\end{center}
\caption{(Color online)
The simulation data for $p_x(v_x)$ and $p_y(v_y)$ from the FRC model and the
XY model are compared with their theoretical estimates. Simulations
and calculations were done at $t=2$ for a polymer with one end grafted in $x$-
direction. 
}
\label{pxpy}
\end{figure}

\begin{figure}[t]
\begin{center}
\includegraphics[width=8.6cm]{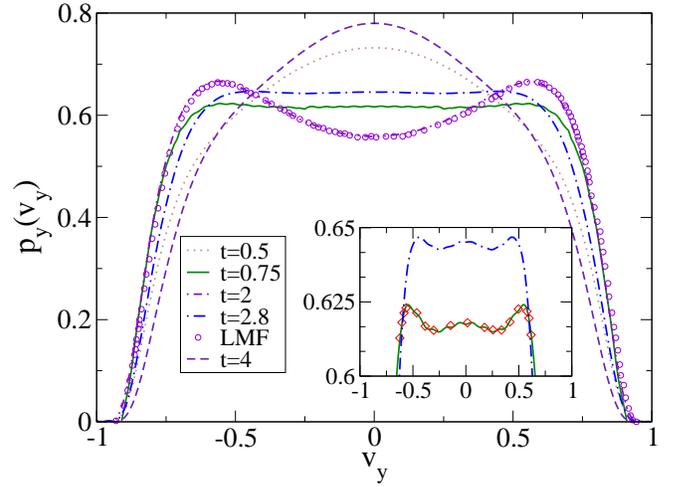}
\end{center}
\caption{(Color online)
For a polymer with one end grafted in $x$- direction,
the integrated probability distribution $p_y(v_y)$ is plotted at various
stiffnesses $t$. At $t=4$ there is a single maximum at $v_y=0$.
Decreasing $t$  we see at $t=2.8$ emergence of two more 
peaks at nonzero $v_y$ apart from the one at $v_y=0$ (See inset). 
At $t=2$ the central peak vanishes, the trimodal distribution becomes bimodal. 
The circles labeled LMF are data taken from Ref.\cite{munk-frey} at $t=2$
and show excellent agreement with our theory.
At $t=0.75$ we see re-emergence of the central peak and trimodality in
$p_y(v_y)$ (See inset, $\Diamond$s are from our MC simulation in the FRC 
model at $t=0.75$.) The lines are calculated from theory. 
}
\label{manyt-py}
\end{figure}

\subsection{Grafted polymer: One end} 

\begin{figure}[t]
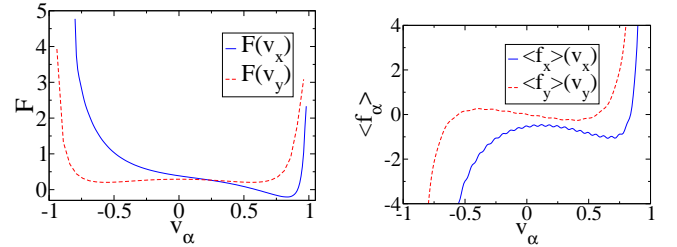

\begin{center}
\includegraphics[width=4.cm]{t2-Feng-teth.eps}
\hskip .4cm
\includegraphics[width=4.cm]{t2-f-constext-teth.eps}
\end{center}
\caption{(Color online)
The left panel shows the Helmholtz free energies ${\cal F}(v_x)$ and
${\cal F}(v_y)$ of a polymer at $t=2$ and one end grafted in $x$- direction.
The right panel shows the corresponding force-extensions in the Helmholtz 
ensemble. Both $\la f_x\ra$- $v_x$ and $\la f_y\ra$- $v_y$ show non- 
monotonicity and regions of 
negative slope. Free energies are expressed in units of $\kb T$ and forces
are expressed in units of $\kb T/\l$.
}
\label{cext}
\end{figure}

\begin{figure}[t]
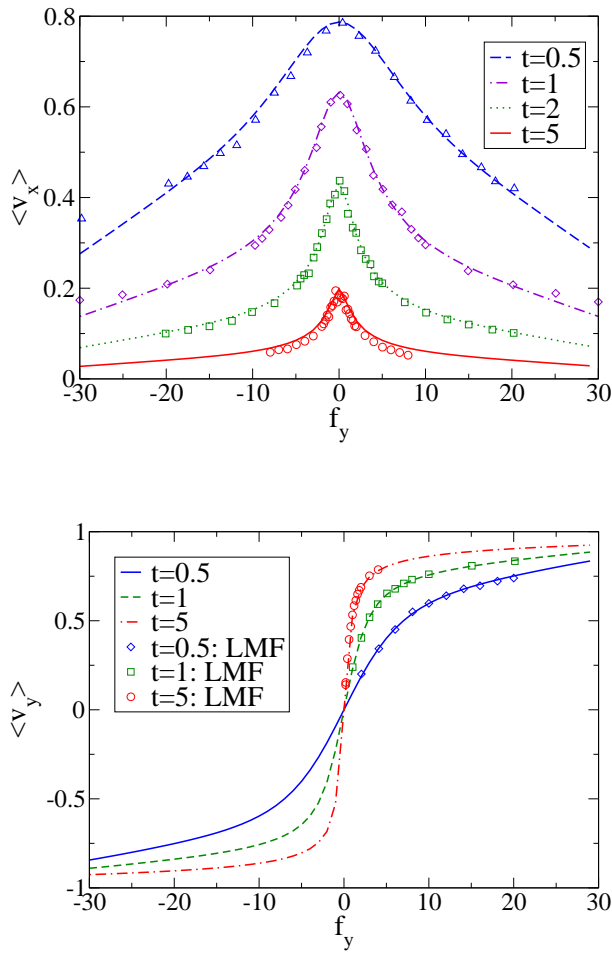

\begin{center}
\includegraphics[width=8.cm]{f-xavg.eps}
\vskip 1cm
\includegraphics[width=8.cm]{f-yavg.eps}
\end{center}
\caption{(Color online)
Average displacements along $x$- direction $\la v_x\ra$ and $y$-
direction $\la v_y\ra$ as a function of transverse force (transverse to 
grafting direction $x$) in constant force ensemble. Lines denote our theoretical
calculation while points denote the MC simulation data taken from
Ref.\cite{munk-frey}. Forces ($F_y$) are expressed in units of $\kb T/\l$, i.e.
$f_y=F_y\l/\kb T$.
}
\label{fx-teth}
\end{figure}

{\em The Helmholtz ensemble:}
Let us compare our theoretical and simulation estimate of
 $p_x(v_x)$ and $p_y(v_y)$ at $t=2$ 
(Fig.\ref{pxpy}) for a semiflexible polymer with one end grafted 
in $x$- direction. 
The excellent agreement validates both our theory 
and the simulation techniques. In $p_x(v_x)$,
the peak in near complete extension along positive $x$ is due to
the coupling of the end orientation towards this direction with
large bending energy (also see Fig.\ref{comp3}).
We then explore, in detail, the transverse fluctuation $p_y(v_y)$ of this 
system for different $t$ (Fig.\ref{manyt-py}).
At large $t(=10)$, $p_y(v_y)$ has single maximum at 
$v_y=0$. At such low stiffnesses entropy takes over energy contributions.
Number of possible configurations and thus entropy gains if end to
end separation remains close to zero. This gives rise to the single
central maximum.  The emergence of multiple maxima at nonzero $v_y$,
the multimodality, at larger stiffness ($t=2.8$) is 
due to the entropy- energy competition. The central peak is due to the
entropy driven Gaussian behavior. The other two peaks emerge as 
entropy tries to fold the polymer and energy restricts the amount
of bending. Since bending in positive and negative $y$- directions are 
equally likely, the transverse fluctuation shows two new maxima near 
$v_y=\pm0.5$ symmetrically positioned around $v_y=0$. 
With further increase in stiffness ($t=2$), 
the central entropic peak vanishes (also see Fig.\ref{comp3}) and 
$p_y(v_y)$ becomes bimodal with two maxima (Fig.\ref{manyt-py}). 
At even higher stiffness ($t=0.75$) the
central peak reappears, due to a higher bending energy.
At $t=0.5$ the distribution again becomes
single peaked at $v_y=0$ as bending energy takes over entropy and
the polymer becomes more like a rigid rod. However, even at 
very high stiffness like $t=0.5$ the single peaked distribution
$p_y(v_y)$ is quite broad underlining the influence of entropic
fluctuations.
Notice that we have plotted MC data taken from Ref.\cite{munk-frey}
for the XY model simulation at $t=2$ (Fig.\ref{manyt-py}). 
This shows very good agreement with our theory. Infact all the simulated
data from Ref.\cite{munk-frey} at different $t$ show excellent agreement
with our theoretical predictions.
In the inset of Fig.\ref{manyt-py}, we have magnified the multimodality
at $t=2.8$ and $t=0.75$. We have also plotted our FRC model simulation 
data at $t=0.75$ and obtained very good agreement.  

At this point, it is instructive to look at the force extension behavior  
in the Helmholtz ensemble, the ensemble in which 
$p_y(v_y)$ and $p_x(v_x)$ have been calculated above. In it the extension $v_x$
[$v_y$] is held constant and the corresponding average force in $x$- 
[$y$-] direction is found from the 
relation $\la f_x\ra = \p {\cal F}(v_x)/\p v_x$
(~or $\la f_y\ra = \p {\cal F}(v_y)/\p v_y$~). 
Notice that, when $v_x$ [$v_y$] is held constant, $v_y$ [$v_x$]
remains free. This can be achieved using a trapping potential constant
in $v_y$ [$v_x$] and trapping the polymer end in $v_x$ [$v_y$].
In Fig.\ref{cext}, we
show the Helmholtz free energies ${\cal F}(v_x) = -(1/t)\ln~p_x(v_x)$ and
${\cal F}(v_y) = -(1/t)\ln~p_y(v_y)$ and the corresponding force extension 
curves in constant extension ensemble. 
Note that unlike the monotonicity obtained in
$\la v_y\ra$-$f_y$ curve (Fig.\ref{fx-teth}) in the Gibbs ensemble,
the $\la f_y\ra$-$v_y$ curve in Fig.\ref{cext} clearly shows  non-monotonicity,
a signature of semiflexibility in the Helmholtz ensemble.

{\em The Gibbs ensemble:}
From our theory we can also explore the transverse response 
of a polymer which has one of its ends grafted and a constant force is 
applied to the other end in a direction transverse to the grafting direction. 
Assume that the grafting direction is $x$ and a force $f_y$ is applied in 
$y$- direction to study the transverse response. A linear response theory 
was proposed earlier\cite{kroy} to tackle this question. 
Our theory can predict the effect of externally applied
force $f_y$ of arbitrary magnitude on the average positions $\la v_x\ra$ and 
$\la v_y\ra$.
As the force is applied in $y$-direction i.e. $\vec f=\hat y f_y$, 
we have $H_I = -f_y \sin\h$. Because one end of the polymer is 
grafted in $x$- direction we use
$\langle n|\hat H_I| n'\rangle = -(f_y/2i)(\d_{n',n+1} - \d_{n',n-1})$
to evaluate $\tilde Z(f_y)$, whereas to calculate 
$\la v_x\ra = -(\p G/\p f_x)$ 
[~or, $\la v_y\ra = -(\p G/\p f_y)$~], we introduce a small 
perturbing force $\d f_x$ [~or, $\d f_y$~] in the Hamiltonian matrix
to obtain the partial derivatives.  Thus we obtain
the corresponding force-extensions shown in Fig.\ref{fx-teth}.
As the grafted end is oriented in $x$- direction, we expect, 
in absence of any external force, $\la v_x\ra$ will be maximum 
and will keep on reducing due to the bending of the other end 
generated by the external force $f_y$ imposed in $y$- direction. Thus 
$\la v_x\ra$ is expected to be independent of the sign of $f_y$. Similarly,
$\la v_y\ra$ should follow the direction of external force and therefore
is expected to carry the same sign as $f_y$.
Fig.\ref{fx-teth} verifies these expectations and
shows very good agreement between our theory and simulated data taken
from Ref.\cite{munk-frey}. It is interesting to note that, in the Helmholtz
ensemble, the multimodality in probability distribution predicts 
non-monotonicity in force-extension relation. However, as expected, this 
non-monotonicity does not survive in the Gibbs ensemble.

\begin{figure}[t]
\begin{center}
\includegraphics[width=8.cm]{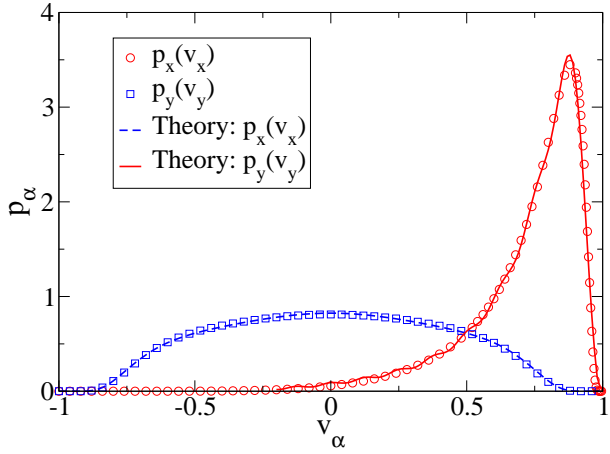}
\end{center}
\caption{(Color online)
The simulation data for $p_x(v_x)$ and $p_y(v_y)$ from the
XY model simulations of a WLC polymer are compared with their theoretical 
estimates. Simulations and calculations were done at $t=2$ for a WLC polymer 
with both its ends grafted in $x$- direction. 
}
\label{both-pxpy}
\end{figure}

\begin{figure}[t]
\begin{center}
\includegraphics[width=8cm]{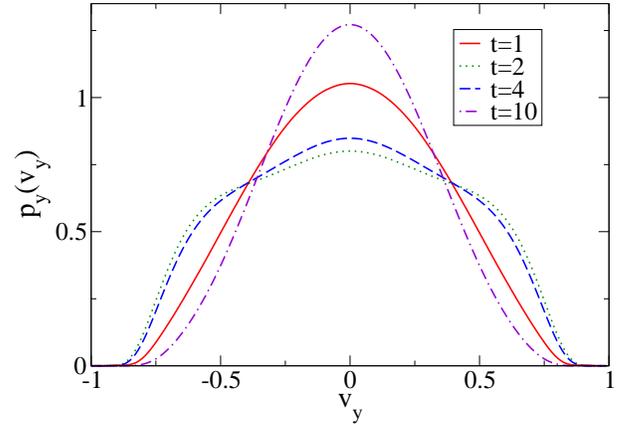}
\vskip 1cm
\includegraphics[width=8cm]{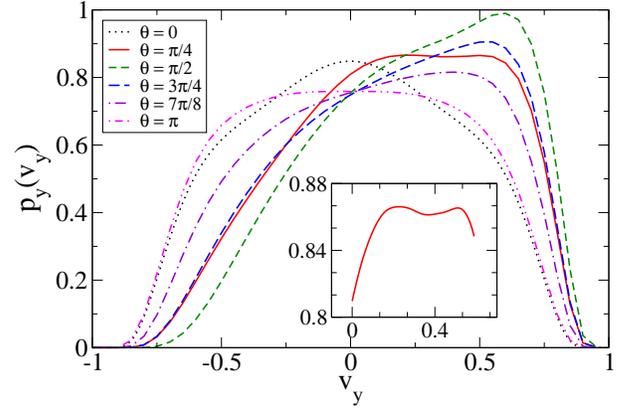}
\end{center}
\caption{(Color online)
The upper panel shows $p_y(v_y)$ for a polymer with both 
ends grafted along $x$- direction at various stiffness parameters $t$.
They always show single maximum. In lower panel, $p_y(v_y)$ is plotted for 
various relative angles $\h$ between the orientations of the two ends  at $t=4$. 
The inset magnifies the emergence of bimodality at $\h=\pi/4$. 
}
\label{bothend1}
\end{figure}
\subsection{Grafted polymer: Both ends}
{\em The Helmholtz ensemble:}
Let us first fix the orientations of the polymer at 
both its ends along $x$- axis and 
compare $p_x(v_x)$ and $p_y(v_y)$ obtained from our XY model simulation
and our theory (Fig.\ref{both-pxpy}). The very good agreement
validates both our theory and simulation.
Then, we go on to explore the properties of this system using the 
theory developed in Sec.\ref{pathint}-C.
Let us fix the orientation at one end in $x$- direction ($\h_i=0$) 
and that in the other end ($\h_f$) can be
varied to study the change in transverse fluctuation $p_y(v_y)$.
To begin with, let us find $p_y(v_y)$ for different stiffness parameters
$t$ with $\h_f=0$ (Fig.\ref{bothend1}). The height of the central peak shows
non-monotonicity -- with increase in $t$ from $t=1$ the
height of the central peak first decreases up to $t=2$ 
and then eventually it increases again. 
The initial decrease in peak height is due to the 
fact that with increase in $t$, i.e. with lowering in stiffness,
 the other end of the polymer (relative to the
first end) starts to sweep larger distances from the $x$- axis. With further
increase in $t$ ($t=4$), the height of the maximum increases
(also see Fig.\ref{comp3}). From Fig.\ref{comp3}, notice that at
$t=4$ multimodality appears in the distribution of end to end vector.
The new entropic maximum at $\vv=\vec 0$ contributes towards increasing
the peak height in $p_y(v_y)$ at $v_y=0$. 
Though, in $p(v_x,v_y)$ multimodality is present
(see Fig.\ref{comp3}) at $t=4$, after integration over probability
weights along $x$- direction the projected distribution $p_y(v_y)$
becomes unimodal. Thus multimodality in the probability distribution
of end to end vector does not guarantee multimodality in projected
probability distributions.
\begin{figure}[t]
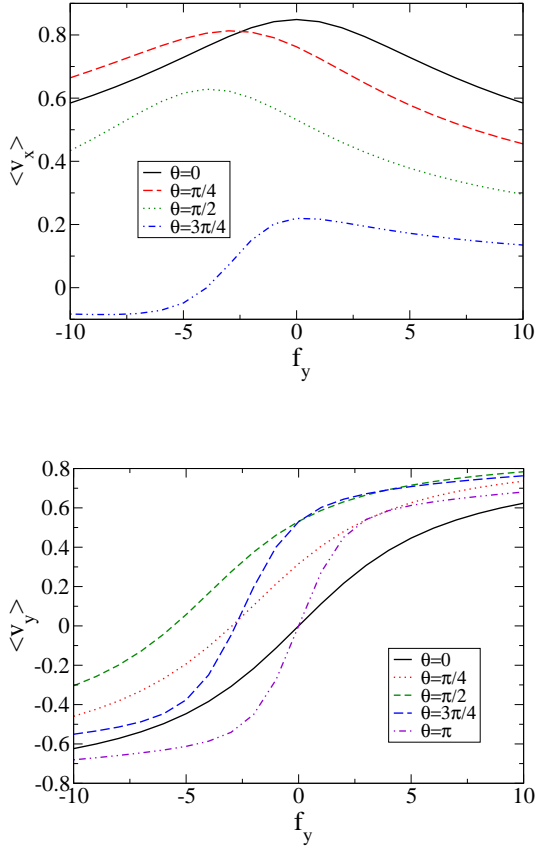

\begin{center}
\includegraphics[width=7cm]{fy-vx-manytheta-t1.eps}
\vskip 1cm
\includegraphics[width=7cm]{fy-vy-manytheta-t1.eps}
\end{center}
\caption{(Color online)
Average displacements $\la v_x\ra$ (upper panel) and
$\la v_y\ra$ (lower panel) as a function of a force $f_y$ for a polymer
having one end grafted along $x$- direction and the other in an angle
$\h$ to the $x$- direction. Forces are expressed in units of $\kb T/\l$.
All the force-extension curves are obtained at $t=1$. 
}
\label{bothend2}
\end{figure}
\begin{figure}[t]
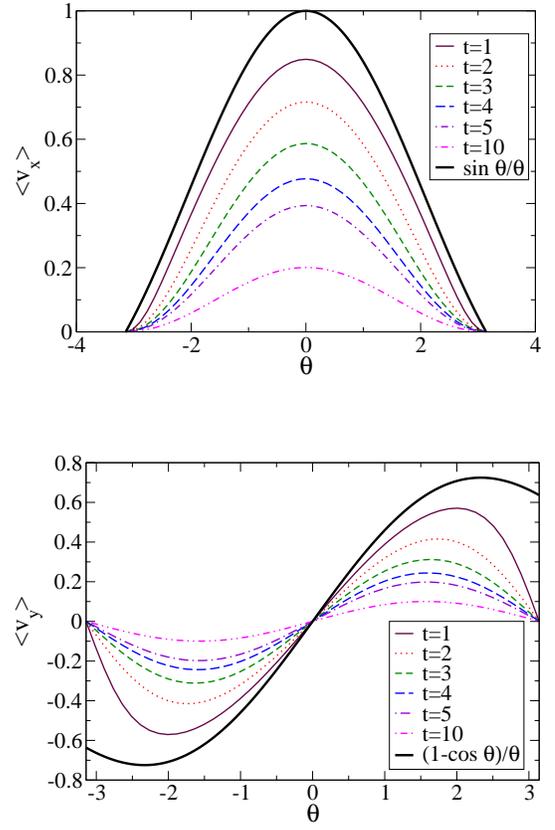

\begin{center}
\includegraphics[width=7cm]{bothend-theta-xavg.eps}
\vskip 1cm
\includegraphics[width=7cm]{bothend-theta-yavg.eps}
\end{center}
\caption{(Color online)
The upper panel shows the variation of $\la v_x\ra$ as a function
of $\h$ and the lower  panel shows the variation of $\la v_y\ra$ as a function
of $\h$. $\la v_x\ra$ and $\la v_y\ra$ are calculated for stiffness
parameters $t=1,2,3,4,5,10$. The thick solid line, in both the plots, show
the expected behavior coming from energetics ignoring the entropy.
}
\label{bothend3}
\end{figure}

To see the impact of change in relative angle of grafting,
we fix one end along $x$- axis and rotate the orientation of the other
end and find out the transverse fluctuation $p_y(v_y)$ at $t=4$
(Fig.\ref{bothend1}). 
At $\h\equiv\h_f=0$ the fluctuation is unimodal with the maximum at $v_y=0$.
With increase in $\h$ the orientation of the other end rotates from
positive $x$- axis towards positive $y$- axis. 
Energetically the polymer gains the most, if it bends along the perimeter
of a circle. 
Therefore,  energetically, at any $\h$, the peak of
$p_y(v_y)$ would like to be at $v_y = (~1-\cos\h~)/\h$.
Thus at $\h=0,\pi/4,\pi/2,3\pi/4,7\pi/8,\pi$ the peak of
$p_y(v_y)$ should be at $v_y = 0,0.37,0.64,0.72,0.69,0.64$ respectively.
Fig.\ref{bothend1} shows that the peak positions almost follow these values
up to $\h=\pi/2$, above which entropic contributions dominate to bring down 
the peak positions to lesser $v_y$ with respect to that attained at $\h=\pi/2$. 
However, entropy always play a crucial role, e.g. at $\h=\pi/4$, 
$p_y(v_y)$ shows a double peak around $v_y = 0.37$.
At $\h=\pi$ the two ends of the polymer are kept anti- parallel.  
Notice that, as $\h=\pi$ and $\h=-\pi$ are physically same,  
at $\h=\pi$, energetically, $v_y=\pm 0.64$ are equally likely. 
Entropy would like the two ends to bend to $v_y=0$. Competition 
between energy and entropy leads to almost a constant 
distribution up to $|v_y|\sim 0.5$. 
The behavior of $p_y(v_y)$
for $-\pi\le \h\le 0$ is mirror symmetric about $v_y=0$ with respect
to the behavior of $p_y(v_y)$ in the region $0\le \h\le \pi$.

{\em The Gibbs ensemble:}
We then work in the constant force ensemble by applying a force 
$\vec f=\hat y f_y$ on an end oriented along any direction $\h$ to $x$-axis
while the other end is oriented along $x$- direction.
We find out the corresponding responses, $\la v_x \ra$-$f_y$ 
and $\la v_y \ra$-$f_y$ to this force (Fig.\ref{bothend2}) 
in the similar manner as has been done in the last subsection for 
the case of a polymer with one end grafted.
If $\h=0$, the force extensions carry the same qualitative 
features as for a single end grafted polymer at all $t$
(see $\h=0$ curves for $t=1$ in Fig.\ref{bothend2}). 
Therefore, instead of showing the $t$ dependence of force-extension
behavior, we show the $\h$ dependence of force
extensions at $t=1$. The peak in $\la v_x \ra$-$f_y$ curve shifts to
$f_y<0$ as $\h$ is increased up to $\pi/2$ above which it again shifts
back towards $f_y=0$. With increase in $\h$, $\la v_x\ra$ decreases,
as with these boundary orientations the polymer is 
forced to close in $x$- and open up in $y$- direction. However, for
$\h\to\pm \pi$ entropy likes $\la v_y\ra\to 0$. For $\h<\pi/2$
small negative $f_y$ leads to unfolding thereby increasing $\la v_x\ra$.
Whereas for $\h>\pi/2$ the effect of negative force is opposite --
it helps the polymer to get folded to reduce $\la v_x\ra$. At 
$\h=\pi$, $\la v_x\ra$ always remains zero.
The responses for negative $\h$ are reflection symmetric about $f_y=0$.
The folding behavior is also apparent from $\la v_y \ra$-$f_y$ curves.
Up to $\h=\pi/2$ the response shifts towards positive $\la v_y \ra$
as the polymer likes to open up in $y$- direction due to the bending
energy cost. However, for large $\h$ entropy wins and 
at $\h=\pi$, $\la v_y \ra$-$f_y$ curve, again, goes through origin. 
The elastic constant $\p f_y/\p \la v_y\ra$ near $f_y=0$ (linear response) 
is larger at $\h=0$ as compared to at $\h=\pi$; i.e. the transverse 
response of a semiflexible polymer with parallel end orientations is
more rigid than with anti-parallel end orientations.
To see the impact of the change in relative angle $\h$, in detail, 
we calculate $\la v_x \ra$ and $\la v_y \ra$ as we vary $\h$ 
(Fig.\ref{bothend3}) keeping external force at zero. 
Bending energy would like $\la v_x\ra=\sin\h/\h$ and 
$\la v_y\ra=(1-\cos\h)/\h$.
Note that at $\h\to 0$, energetically, $\la v_x\ra\to 1$ and 
$\la v_y \ra \to \h/2$.
Again, at $\h\to\pm\pi$ bending energy requires
$\la v_x\ra\to 0$ and $\la v_y\ra\to \pm 2/\pi$ though
entropy likes $\la \vv\ra\to \vec 0$. Thus at small $t$, the approach of
$\la v_x\ra$-$\h$ curve to $\sin\h/\h$ is much better than 
approach of $\la v_y\ra$-$\h$ to $(1-\cos\h)/\h$ (Fig.\ref{bothend3}).
It should be noted that the angle $\h$ in this study denotes a relative
angle of bending between the two end orientations of a WLC polymer.
This should not be confused with the twist angle as in Ref.\cite{twist-2}.
In an earlier study\cite{wang} the impact of changing $\h$
on the averaged root mean squared end to end vector
has been obtained. In this section we have shown the impact of changing
$\h$ on projected probability distribution, averaged end to end distance
($\la v_x\ra$, $\la v_y\ra$) and force-extension relations.

\subsection{Distribution of end to end vector }
\begin{figure}[t]
\begin{center}
\includegraphics[width=8.0cm]{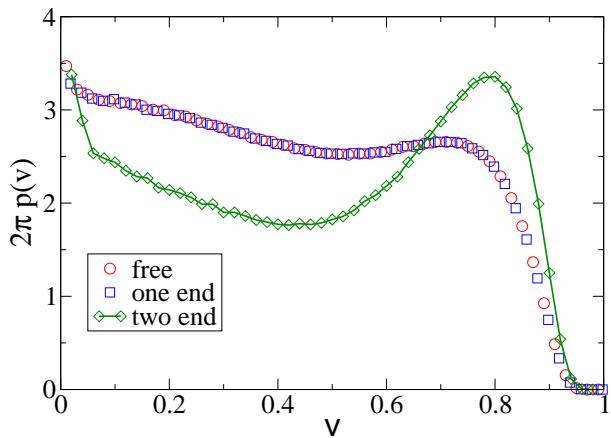}
\end{center}
\caption{(Color online)
The probability distribution of end to end distance $2\pi p(v)$ at 
stiffness $t=4$ is plotted for the three different boundary conditions -- 
(a) both ends free, 
(b) one end oriented in $x$- direction and the other kept free, 
(c) both ends oriented in $x$- direction. Radial distribution of 
first two cases are equal, whereas for the third case it is different.
However, all the three curves show double maxima.
}
\label{comp1}
\end{figure}

\begin{figure}[t]
\begin{center}
\includegraphics[width=8cm]{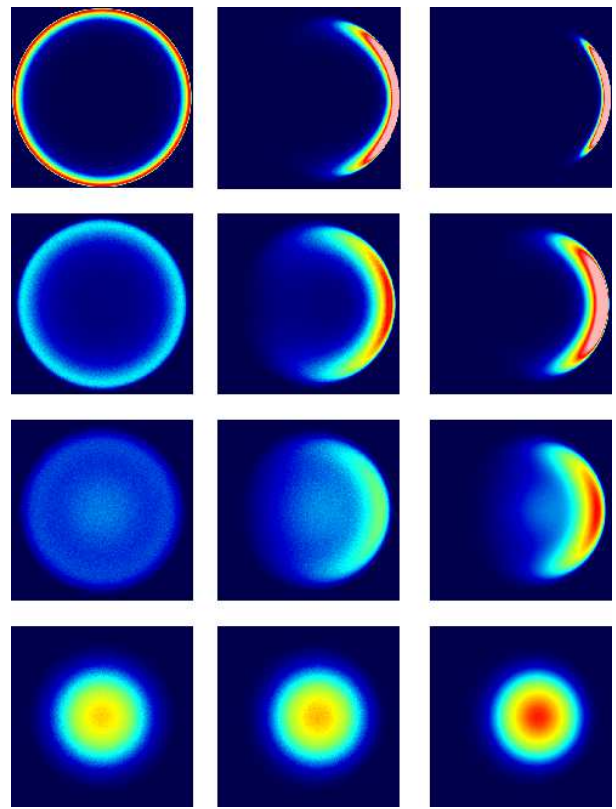}
\end{center}
\caption{(Color online)
Density plot of $p(v_x,v_y)$. Color code: red (light) - high density,
blue (dark) - low density. Left panels are for free polymers, middle
panels are for polymers having one end grafted in $x$- direction and the 
right panels are for polymers having both ends grafted in $x$- direction.
From top to bottom four panels denote increasing stiffness parameters 
$t=0.5,2,4,10$ (decreasing stiffness).
Note that the double maxima feature in $p(v_x,v_y)$ 
(one maximum near the centre and another near the rim) 
at $t=4$ persists for all the three boundary conditions.
}
\label{comp3}
\end{figure}

We now employ MC simulations to study some other aspects of
probability distribution. We first examine the probability distribution 
of end to end distance $p(v)$. It
is clear from Fig.\ref{comp1} that grafting one end does not change the
double maxima feature in $p(v)$ at intermediate values of stiffnesses 
($4\le t\le2$). This is because the two cases are symmetry related; 
fixing orientation at one end only shifts
the probability weight distributed over all possible angles at a given
radial distance $v$ towards the direction of the orientation.
Though grafting both the ends change the distribution of end to end distance,
the double maxima feature persists and becomes more pronounced.
We note that once one end of a polymer is grafted immediately the system
loses its spherical symmetry, more so, since we restrict ourselves to 
semiflexible regime. For a free polymer $p(v)$ plays the role of a
probability distribution of end to end vector and thus gives the Helmholtz
free energy and force-extension behavior. Once the spherical symmetry is 
broken $p(v)$ merely plays the role of a radial distribution function 
in terms of $2\pi v p(v)$ and no longer remains relevant in
predicting the force-extension behavior.
We have already seen that the projected probability distributions
$p_x(v_x)$ and $p_y(v_y)$ are very different for grafted polymers,
though they are the same for free polymers that preserve spherical
symmetry.

The full statistics of the WLC polymers are encoded in 
the end to end vector distribution function 
$p(v_x,v_y)$. To see the complete structure, we next 
obtain $p(v_x,v_y)$ from MC simulations in the FRC (for a free polymer
or a polymer with one end grafted) and the XY model (for a polymer 
with both ends grafted) and present them as two dimensional density plots. 
We compare $p(v_x,v_y)$ of a free polymer, a polymer with one end  grafted 
and a polymer with both ends grafted (Fig.\ref{comp3}). For definiteness, we
chose all the graftings, fixing of end orientations, 
to be in the $x$- direction.
We plot $p(v_x,v_y)$ over a range of stiffnesses ($t=0.5,2,4,10$). 
The distribution has finite values for $v\leq 1$ and is zero for 
$v>1$. This is due to the inextensibility constraint in the WLC model.
In these density plots high probability is shown in red (light) and low 
in blue (dark) (Fig.\ref{comp3}). At small stiffness ($t=10$) $p(v_x,v_y)$
shows a single entropic peak at $\vv=\vec 0$ for free polymer 
(Fig.\ref{comp3}). This is slightly
shifted towards the direction of end- orientations in grafted polymers.
This shifted entropic peak slowly moves towards $\vv=\vec 0$ in the 
$t\to\infty$ limit. 
With increase in stiffness ($t=4$), a new energy dominated probability peak 
appears near the full extension limit, $v=1$, of the polymer (Fig.\ref{comp3}). This peak forms a circular ring for free polymers. For a grafted polymer, 
this new peak is aligned in the direction of grafting. The probability 
distribution $p(v_x,v_y)$ at $t=4$ clearly shows two regions of probability 
maxima, one near the zero extension and another near the full extension, 
for polymers with all kinds of boundary orientations -- the free polymer, 
the polymer with one end grafted and the polymer with both ends grafted.
Grafting of polymer ends, in a sense, enhances the effective stiffness.
Therefore with increase in polymer stiffness (decrease in $t$) the 
multimodality sets in first in the polymer with both the ends grafted
in the same direction near $t=6$. 
At this $t$ value the free polymer and the polymer with one 
end grafted show only the entropic peaks near $\vv=\vec 0$. Near $t=5$, the
polymer with one end grafted starts to show multimodality. For free polymers,
multimodality sets in only at an even higher stiffness close to $t=4$.
These behaviors are also borne out by the theory.
The multimodality (two maxima) in probability distribution of end to 
end vector seen for a free polymer at $t=4$ (Fig.\ref{comp3}) 
gives rise to the triple minima in free energy found in Ref.\cite{mywlc}. 
In an earlier work\cite{wang} it was shown that the crossover from 
flexible chain to rigid rod via multimodality in probability distribution
as obtained for a free polymer\cite{mywlc} persists even after
grafting one end of the polymer. Here we have shown that this behavior
persists even after grafting both the ends of a polymer. Certainly
the detailed features of probability distribution of end to end vector
would change with changing the relative angle of grafting at the two ends.
At an even larger stiffness ($t=2$), the entropic maximum near the centre 
($\vv=0$) disappears (Fig.\ref{comp3}). For the free polymer, one energy
 dominated maximum gets equally distributed over all angles. 
This way the system uses its spherical symmetry 
to gain in entropy. For grafted polymers, probability maximum near the
full extension fans a finite solid angle around the direction of grafting.
The distribution around the grafting direction is narrower for the polymer
with both its ends grafted along the same direction. 
This is due to a larger coupling between grafting and bending stiffness.
This fact is more pronounced in $p(v_x,v_y)$ at $t=0.5$ (Fig.\ref{comp3}).

\begin{figure}[t]
\begin{center}
\includegraphics[width=8cm]{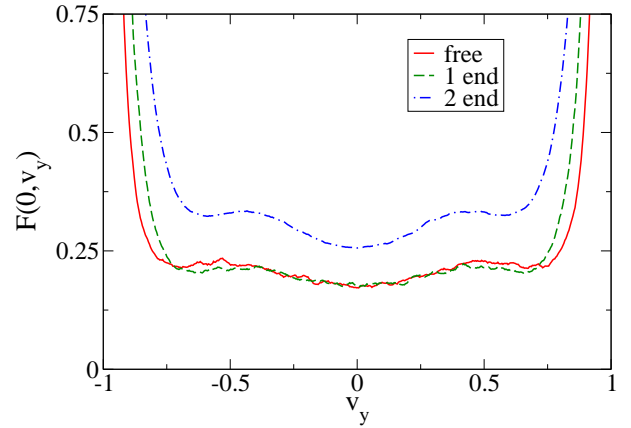}
\end{center}
\caption{(Color online)
At $t=4$ free energy profile ${\cal F}(0,v_y)$  
corresponding to the probability distributions shown in Fig.\ref{comp3}
are plotted.
This clearly shows that the triple minima feature in free energy for
a polymer with both ends free persists
even after grafting one or both ends of the polymer. 
}
\label{comp4}
\end{figure}
\begin{figure}[t]
\begin{center}
\includegraphics[width=8cm]{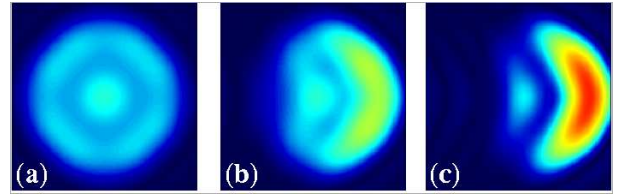}
\end{center}
\caption{(Color online) Density plot of  $p(v_x,v_y)$ obtained from theoretical 
calculations. 
Color code: red (light) means high probability and blue (dark) means low. 
All the plots are obtained for stiffness $t=4$; (a) ends free to rotate, 
(b) one end grafted along $x$- direction,
(c) both ends grafted along $x$- direction.
}
\label{th-chand}
\end{figure}

As mentioned earlier, 
in the Helmholtz ensemble the free energy is given by 
${\cal F}(v_x,v_y) = -(1/t)~\ln [p(v_x,v_y)]$.
This free energy will give the force-extension behavior if the ends are
trapped in 2D plane at some points $(0,0)$ and ($v_x,v_y$).
In Fig.\ref{comp4} we plot this free energy profile ${\cal F}(0,v_y)$
at $t=4$ and compare the three different boundary conditions. 
This plot clearly shows that triple minima in free energy \cite{mywlc} 
prevails even after grafting one or both ends of a semiflexible polymer. 
In terms of force-extension what this triple minima means? If we start off
with end to end vector at ($0,0$) and increase $|v_y|$, for small extensions
the ends would experience an attractive force between themselves. Beyond
a limit ($|v_y|\stackrel{>}{\sim}0.5$) the ends would repel each other to 
take the system to the other minima at non-zero $v_y$. 
At very large extension, again, they would experience an attractive 
force, governed by the inextensibility constraint.
Thus, in force-extension
experiments on a polymer in constant extension ensemble, this multistability
(non-monotonic force extension) at intermediate stiffness values should be 
measurable for all kind of boundary conditions. However the 
measurement would require averaging over a large number of 
observations as indicated in Ref.\cite{mywlc}. 
At this point, it is interesting to notice that, for a polymer with one
end grafted along $x$-direction the force-extension obtained from the slope
of ${\cal F}(v_x=0,v_y)$-$v_y$ curve gives the transverse response in constant
extension ensemble when the other end is constrained to be at a fixed 
$\vv=(0,v_y)$.
The behavior of the transverse response, evidently, would
then also depend on the fixed value of $v_x$ at which one measures the
response. This is in contrast to the measurement of transverse response by
trapping the polymer end at a constant $v_y$, while leaving it free to move
in $x$-direction. Thus we reemphasize that the force-extension behavior
depends on the kind of trapping potential used in an experiment.
Apart from this, as we have shown, the orientational boundary conditions at
the ends of a polymer and the ensemble of experiment will affect 
the force-extension behavior non-trivially.

Recently, using a Greens function calculation of a WLC polymer with 
one of its ends grafted, the presence of multiple maxima in $p(v_x,v_y)$ 
has been observed\cite{wang}.
In this subsection, we have used MC simulations 
to study $p(v_x,v_y)$ for all the possible boundary conditions. 
We have shown that multiple maxima in $p(v_x,v_y)$ persists near 
$t=4$ for all the three different boundary orientations.
We now utilize our theoretical methods as developed in Sec.\ref{pathint}
to obtain the density plot
for $p(v_x,v_y)$ at $t=4$ for all three different end orientations 
(see Fig.\ref{th-chand}). 
The results plotted in Fig.\ref{th-chand} using $N_d=5$ already shows
good agreement with Fig.\ref{comp3}. This clearly brings forth the presence
of multimodality in the Helmholtz ensemble.
Increase in the number of basis states $N_d$ 
(infinite in principle) will lead to better agreement.

\begin{figure}[t]
\begin{center}
\includegraphics[width=8cm]{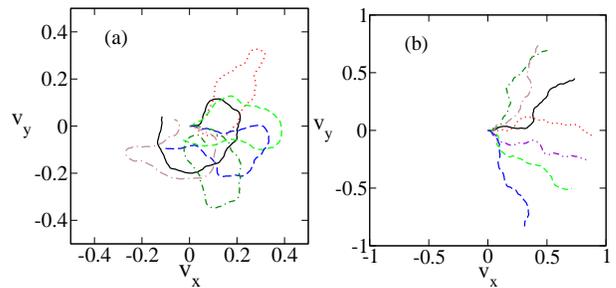}
\end{center}
\caption{(Color online)  Representative conformations at $t=4$,
for polymers with one end grafted in $x$-direction, are shown. 
At this stiffness value the conformations are either
localised corresponding to Gaussian chain like behaviour (a),  or extended 
corresponding to rigid rod like behaviour (b).
}
\label{conforms}
\end{figure}

Is it possible to test the results presented here experimentally?
Fluoroscence microscopy of cortically confined actin filaments has been
performed to extract their persistence length\cite{ott}.  
While most of the force extension measurements are usually done in 3D,
however, experiments on cortical actin filaments are, we believe, possible and 
one may indeed test the predictions described above in such systems.
In this context,
it is instructive to note that, to achieve the parameter
regime $t\sim 4$, for instance, in actin polymers that have persistence length 
$16.7\mu m$ one requires contour length $L\sim 67 \mu m$ which is easily 
achievable experimentally (filaments as long as 100$\mu$m have been 
reported \cite{ott}). 
$L$ can be changed chemically by the addition of enzymes. To measure
the multimodality predicted in this paper, one can perform direct video 
microscopy of the conformations of actin polymers confined in a cell of 
depth $\sim 1\mu m$, practically restricting all fluctuations in third 
direction making the embedding space 2D as in Ref.\cite{ott}. 
In such a setup one can also attach one of the ends 
of the actin molecules to one of the confining glass walls of the
cell that contains them. Thus a setup as in Ref.\cite{ott} may be used
to obtain the probability distribution of end to end vector for a free polymer
as well as a polymer with one end grafted. 
With actins of contour length $\sim 67 \mu m$  the probability distribution
of end to end vector should show multimodality implying bistability in the 
force extension measurement in the Helmholtz ensemble. In this stiffness 
regime even the projected probability distribution is expected to show 
multimodality in 2D in contrast to 3D. 
In typical force-extension measurements one or both the ends of a polymer 
are attached
to dielectric or magnetic beads to hold the ends optically or magnetically.
In a recent study\cite{pnelson} it was shown that to extract physically 
meaningful results from such experiments on dsDNA one has to incorporate 
bead geometry explicitly in the theoretical modelling, since the typical
bead radius $R$ is in between $0.05$ -- $0.5\mu m$\cite{pnelson} which is about
one to ten times the persistence length ($\l=50 nm$) of dsDNA. However, since
the persistence length of actin filament is much larger ($\l=16.7\mu m$),
for actin $R/\l$ is in between $0.003$ -- $0.03$, one does not need to
worry about bead geometry in analyzing the force-extension results for
actin filaments. Our theoretical predictions can thus be straightaway tested
in experiments on actin. Note that, semiflexible polymers in 3D are also
expected to show multimodality in the probability distribution of end to end
vector and therefore non-monotonic force-extension in Helmholtz ensemble 
at least with end orientations free to rotate\cite{mywlc}. 
In Fig.\ref{conforms} we show some typical conformations of a semiflexible 
polymer at $t=4$ lying in 2D embedding space. At this 
stiffness, there are two maxima in the end to end separation, one is near 
the zero separation and the other is near the full extension 
(see Figs.\ref{comp3} and \ref{th-chand}).  Fig.\ref{conforms}.(a)
shows some representative conformations with nearly zero extension and 
Fig.\ref{conforms}.(b) shows the same at near full extension.
 
\section{conclusion}

\label{conclusion}
In this paper, we have shown that the results of force-extension
experiments on semiflexible polymers would depend on ensemble, 
constraints on end orientations, dimensionality of embedding space
and the kind of trapping potential used. In an earlier work we
have shown the presence of multiple maxima in the probability distribution
of end to end distance of a free polymer at intermediate stiffnesses 
(near $t\sim 4$) that lead to non-monotonic force-extension in the Helmholtz 
ensemble\cite{mywlc}. In this paper, we have demonstrated that though the 
details of the end to end distribution depends crucially on the constraints 
imposed on the end orientations, the multimodality
in the distribution always persists. In this paper, we have used a mapping of 
the WLC model to a quantum particle on a sphere to obtain probability
distribution of end to end separation and force-extension in various
ensembles taking care of the particular types of the constraints on the
orientations of polymer ends. 
We have made a number of predictions about the end to end statistics
and the force-extension behaviors in the Helmholtz and the Gibbs ensemble.
We have used MC simulations 
against which we have tested the theoretical predictions and always
obtained very good agreement. 
Experiments using the laser trap to hold the ends of a polymer allows
all possible end orientations, whereas the magnetic tweezers can be used to
fix the end orientations and see the impact. 
On the other hand it is possible to obtain video microscopy of semiflexible
polymers like actin to obtain the probability distribution of end to end
separation.  Thus it is possible to test our theoretical predictions in 
experiments. In this work we have restricted ourselves to 2D.
At the onset we have shown that an important feature of polymer statistics,
multimodality in projected distributions, is dependent on the dimensionality 
of embedding space. In three dimensional free polymers,
multimodality in projected probability distribution is impossible,
however presence of this is a reality in 2D. 
We have shown that depending on whether the dimensionality 
$d_r$, in which the trapping potential traps the polymer ends,
 is same or less than the dimensionality of
embedding space $d$, the physically relevant Helmholtz free energy 
would be obtained from the probability distribution of end to end
vector or a $d_r$-dimensional projection (~$[d-d_r]$ dimensional integration~)
of it. 
After projection, multimodality in the distribution 
function of end to end vector may or may not survive, thereby
affecting the qualitative features of the force-extensions.
Fixing the orientation of a WLC polymer at one end we have studied the 
projected probability distributions in the longitudinal and 
transverse directions. The transverse fluctuations and
the force-extensions found from our theory show
excellent agreement with MC simulations in Ref.\cite{munk-frey-th}. 
If orientations at both the ends are
kept fixed, the polymer properties vary depending on the relative 
angle between the two grafted ends. For example, multimodality in
projected distribution depends on the relative angle. The full
statistics of the WLC polymers are encoded in probability distribution
of end to end vector. Our simulations and theory have clearly shown
that, the multiple maxima feature in this probability distribution 
in the intermediate stiffness regime (near $t=4$)
survives the fixing of end orientations.
Similar studies in 3D remains to be an interesting direction forward. 
Multimodality in probability distribution may show multistability
in the time-scale the end to end separation of
a WLC polymer spends in each of the free energy
minima. In polymer looping, the closing time and the opening time 
of the two ends of a free polymer depends on the polymer stiffness.
The impact of the triple minima in the Helmholtz free energy 
on these time- scales of a free polymer remains to be studied.
This might be of importance in understanding the very fast time-scale
of transcription, with respect to the diffusion time, 
in the process of gene expression\cite{cell}.
We intend to report on some of these problems in future.

\acknowledgments

Illuminating discussions with Abhishek Dhar, Surajit Sengupta and 
Gautam I. Menon are gratefully acknowledged. I thank 
Gautam Mukhopadhyay and Swarnali Bandopadhyay for discussions and 
Jayendra N. Bandyopadhyay, Tamoghna Das and Arya Paul for critical comments
on the manuscript. Partial financial support by CSIR, India and 
computational facilities from DST grant SP/S2/M-20/2001 
and S. N. Bose National Centre for Basic Sciences are greatfully acknowledged.


\begin{thebibliography}{10}
\bibitem{matexp}
 We use the MatrixExp function of Mathematica\cite{mathematica}.
\end{thebibliography}

\end{document}